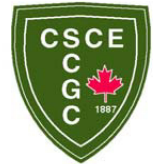

# Supporting Construction Worker Well-Being with a Multi-Agent Conversational AI System

Fan Yang[1]; Yuan Tian[2]; and Jiansong Zhang, PhD., A.M.ASCE[3]

[1]School of Construction Management Technology, Purdue Univ., West Lafayette, IN, USA.
Email: yang2352@purdue.edu
[2]Department of Computer Science, Purdue Univ., West Lafayette, IN, USA.
Email: tian211@purdue.edu
[3]School of Construction Management Technology, Purdue Univ., West Lafayette, IN, USA.
Email: zhan3062@purdue.edu

**ABSTRACT:** The construction industry is characterized by both high physical and psychological risks, yet supports of mental health remain limited. While advancements in artificial intelligence (AI), particularly large language models (LLMs), offer promising solutions, their potential in construction remains largely underexplored. To bridge this gap, we developed a conversational multi-agent system that addresses industry-specific challenges through an AI-driven approach integrated with domain knowledge. In parallel, it fulfills construction workers' basic psychological needs by enabling interactions with multiple agents, each has a distinct persona. This approach ensures that workers receive both practical problem-solving support and social engagement, ultimately contributing to their overall well-being. We evaluate its usability and effectiveness through a within-subjects user study with 12 participants. The results show that our system significantly outperforms the single-agent baseline, achieving improvements of 18% in usability, 40% in self-determination, and 60% in social presence and trust. These findings highlight the promise of LLM-driven AI systems in providing domain-specific support for construction workers.

## 1. INTRODUCTION

The construction industry is characterized by a high incidence of worker injuries and a prevalence of mental health challenges (Langdon and Sawang 2018; Ross et al. 2022). Reports indicate that the suicide rate among construction workers is 44% higher than the national average, underscoring the severity of the problem (Campbell and Gunning 2020). Factors contributing to these challenges include the demanding nature of construction work, job insecurity, and a pervasive culture that often discourages seeking help due to masculinity culture and mental health stigma (Eyllon et al. 2020; Gunning and Cooke 1996; Ness 2012).

In recent years, artificial intelligence (AI) chatbots have emerged as significant tools in mental health support, offering accessible and immediate assistance. These AI-driven platforms provide users with a sense of anonymity, which can reduce the stigma associated with seeking help and encourage open communication. While existing chatbot systems have been implemented in various sectors, there is a noticeable gap in solutions specifically tailored to the unique needs of construction workers. Current general-purpose AI applications often fall short in addressing the specific challenges faced by this group, such as the need for industry-specific safety information and support mechanisms that resonate with their experiences.

In response to the growing need for innovative support systems in the construction industry, we propose a novel multi-agent AI system powered by large language models (LLMs) and retrieval-augmented generation (RAG) technology. Our system incorporates several key innovations: (1) An efficient agent configuration interface that enables rapid customization of agents with distinct conversational goals, personalities, and domain-specific knowledge integration from external documentation; (2) A novel collaborative workflow where specialized agents provide complementary perspectives based on their unique focus areas; and (3) A natural group conversation simulation in which AI agents can dynamically decide whether and when to respond, creating a smoothly supportive system for construction workers.

To evaluate our proposed system, we conducted a controlled experiment comparing it with a single-agent baseline with vanilla conversation settings. This investigation centered on the key research question: *How does the proposed system affect usability, basic psychological needs (competence, autonomy, and relatedness), social presence and trust during the Worker-AI-Interaction rather than generalized LLM?*

## 2. BACKGROUND

### 2.1 Emotional Health Support for Construction Workers

The construction industry is characterized by high physical risks, demanding workloads, and job insecurity, all of which contribute to significant mental health challenges among workers. Studies indicate that construction workers experience elevated rates of stress, anxiety, depression, and even suicidal ideation compared to workers in other industries (Lingard and Rowlinson 2004; Martin et al. 2016). Despite these alarming statistics, mental health support in the construction industry remains limited due to stigma, lack of awareness, and the transient nature of construction employment (Gómez-Salgado et al. 2023).

Interventions have been proposed to address mental health challenges in construction. These include peer support programs, workplace mental health training, and digital mental health interventions (Greiner et al. 2022; Nwaogu et al. 2019). Digital solutions, including mobile applications and online counseling, have emerged as viable methods to provide accessible and scalable support to workers who may be reluctant to seek traditional therapy. However, the effectiveness of these solutions is often constrained by low engagement and the absence of industry-specific tailoring.

### 2.2 Conversational AI in Construction

Conversational AI, encompassing chatbots and virtual assistants, has gained attention as a tool for enhancing communication and information dissemination in construction. AI-driven conversational systems have been employed for various applications, including safety training, project management, and worker assistance. These systems leverage natural language processing (NLP) to facilitate real-time interactions, providing users with relevant information and support without the need for human intervention. However, the overall deployment of conversational AI in the AEC industry is still relatively slow (Saka et al. 2023).

In the domain of worker well-being, conversational AI presents an opportunity to offer personalized and continuous mental health support. AI-powered chatbots designed for mental health applications have demonstrated effectiveness in providing cognitive behavioral therapy (CBT)-based interventions, stress management techniques, and crisis support (Fitzpatrick et al. 2017). However, generic AI chatbots may lack contextual awareness of the unique challenges faced by construction workers. Therefore, the design and development of domain-specific conversational AI tailored to the construction industry, and its potential impact on workers' well-being, is a very important research question.

### 2.3 Multi-agent System

Multi-agent systems (MAS) has gained significant attention for its ability to enhance efficiency, robustness, and adaptability in complex tasks (Dorri et al. 2018). Unlike single-agent approaches, MAS leverages the principle of specialization and division of labor, enabling individual agents to focus on specific subtasks while improving overall performance (McArthur et al. 2007).

The strengths of MAS manifest in several key aspects. First, their parallel processing capabilities enable concurrent task execution, significantly reducing latency and enhancing scalability in real-time applications (Shu et al. 2024). Second, the distributed nature of MAS provides inherent robustness, as the system can maintain functionality even when individual agents fail, ensuring reliable performance in dynamic environments (Chaaban and Müller-Schloer 2013). Third, multi-agent collaboration enhances decision-making accuracy through sophisticated verification and consensus mechanisms, effectively mitigating the biases and errors found in single-agent systems (Amirkhani and Barshooi 2022). Fourth, multi-agent architectures allow for adaptive learning and decentralized control, making them particularly suitable for applications in autonomous systems, financial trading, and human-AI interaction (Rizk et al. 2018).

The emergence of large language models (LLM), such as OpenAI's ChatGPT and Google's Gemini, further facilitate the development of MASs since they exhibit promising capabilities in interpreting natural language instructions. Prior works have applied MASs to several industries, such as chemistry (Boiko et al. 2023). However, there remains a significant research gap in understanding the interaction between human users and LLM-based MAS in conversational scenarios. Our study addresses this gap by developing a novel conversational MAS specifically designed for construction workers. To the best of our knowledge, this represents the first investigation into the application of conversational MAS in the construction domain.

## 3. METHODOLOGY

### 3.1 Overview

We experimented on system deployment, scenario-based interactions, and a user study to evaluate our proposed system. Figure 1 illustrates the proposed RAG-based conversational multi-agent system. User messages are processed by multiple collaborative agents that leverage a vector database and configurable external documentation. Each agent specializes in a distinct domain, such as regulatory explanation or emotional support, and works complementarily to provide multi-faceted responses.

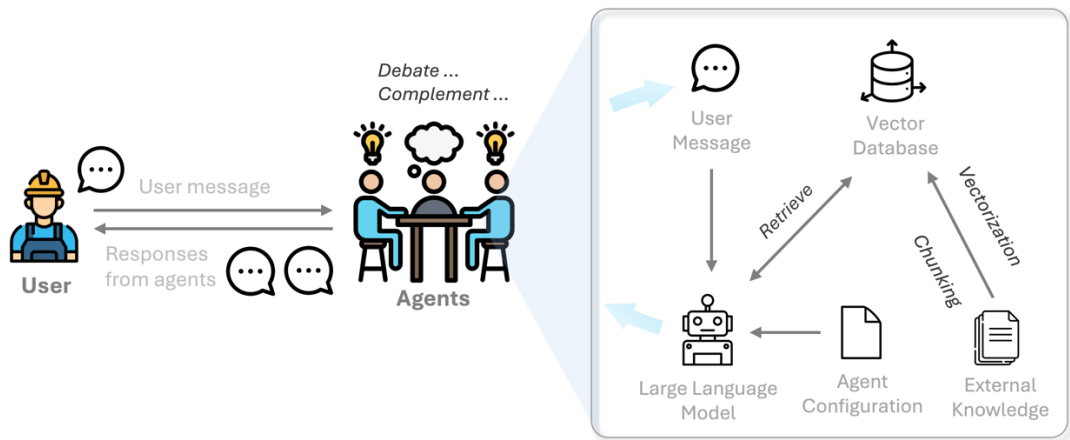

Figure 1: Overview of the proposed conversational multi-agent conversational system

In the following sections, we will first discuss the design of user interface (Section 3.2). Then we discuss details of our system design, including agent configuration (Section 3.3), multi-agent orchestration (Section 3.4). Finally, we evaluate the proposed system by comparing it to the baseline in terms of usability and user experience (Section 3.5).

### 3.2 User Interface Design

Figure 2 demonstrates the user interface (UI) of our proposed system. The UI simulates a group conversation. It is implemented as a React-based web application designed to facilitate natural interactions between the user and multiple AI agents. The UI also features a dynamic agent management panel where users can add or delete active agents (Fig. 2 ①). Each agent is configurable (Fig. 2 ②), allowing users to set the agent's name, personality, conversation goals, and private knowledge, all represented by a customizable avatar. The main conversation area displays message bubbles with clear visual distinctions between user and agent messages (Fig. 2 ③).

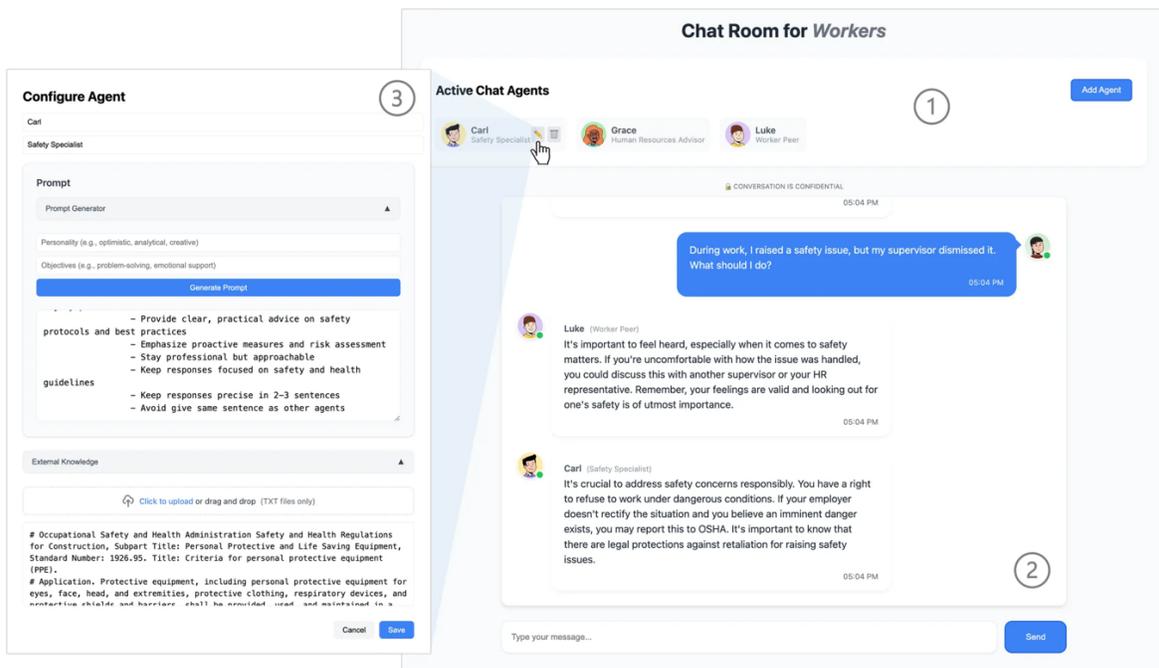

Figure 2: User interface of our proposed system

### 3.3 Agent Configuration

#### 3.3.1 Prompt Engineering and Automation

To ensure consistent agent behavior while reducing the manual effort required for prompt engineering, our system automates the generation of agent descriptions through a two-tier prompt chaining process. Specifically, users can describe the agent using a few keywords, such as the agent's occupation, personality, and conversation goals. Our system then automatically generates the prompt for this agent at different stages, following a consistent format and instructions.

#### 3.3.2 Retrieval-Augmented Generation

Retrieval-Augmented Generation (RAG) is a widely used method that enhances LLMs by first retrieving relevant information from a knowledge base before generating informed responses. To enable knowledge-based discussions, we utilize RAG to enhance each agent with distinct domain-specific knowledge. The system processes documents through chunking and indexing, utilizing a FAISS vector database with OpenAI's semantic embeddings for efficient retrieval. During conversations, the system dynamically retrieves relevant knowledge based on user message and integrate it into the response, while maintaining natural dialogue flow, avoiding explicit references to the knowledge source.

### 3.4 Multi-Agent Orchestration

Our system manages multi-agent interactions through an orchestration system that determines (1) who should respond and (2) the order in which these agents should respond.

Given a user message, the orchestration system enables each agent to evaluate the message relevance and respond only when directly addressed or when their expertise is relevant. For example, in a message like "Hey Alice, what do you think?" only Alice responds, while a general question such as "What do you think of this idea?" might elicit responses from all agents. Our system ensures that at least one agent responds to a user message.

The orchestration system randomly groups agents and performs in either sequential or parallel modes to simulate real-world communication patterns. In sequential mode, agents respond in a randomized order to mimic natural turn-taking. For instance, an agent can reference or give an opinion on another agent's response. Since each agent has a unique role, subsequent agents may support, complement, or completely disagree with preceding agents. In parallel mode, all agents respond simultaneously, offering perspectives from different domains. This feature can significantly reduce response latency.

While each agent's external knowledge remains private, the conversation history of the group is shared among all agents. By accessing this shared history and understanding the roles of other agents, each agent can contribute to the discussion by building on previous responses while maintaining its unique perspective. This design facilitates coherent group discussions that benefit from diverse viewpoints.

**3.5 Experimental Design**

3.5.1 Personas of agents

To conduct comparative experiments within a specific problem domain, we researched the most pressing issues currently facing the construction industry and predefined three representative agents' persona to address the most critical mental health and safety needs of construction workers, each representing a key stakeholder in workplace safety and mental well-being:

> **Occupational Safety and Health (OSH) Specialist Agent**: Provides expert knowledge on workplace injury and illness prevention, promoting proactive safety measures. This agent can access external knowledge covering OSHA (Occupational Safety and Health Administration) regulations, industry safety standards, and compliance guidelines.
>
> **Human Resources Advisor Agent**: Offers insights into workplace safety regulations, employer responsibilities, and employment relationships while promoting a positive workplace culture. This agent can access external knowledge covering OSHA employer obligations, labor laws, workplace mental health programs, return-to-work initiatives, and conflict resolution.
>
> **Worker Peer Agent**: Functions as a social support system, offering empathy, stress management strategies, and peer-based emotional support to create a more open and supportive work environment. This agent can access external knowledge covering stress, burnout coping mechanisms, and guidelines on peer counseling and support systems.

To ensure a fair comparison, we implement the single-agent baseline using the same interface. The only difference is that we replace the backend system with the original AI Chatbot, which lacks predefined personas and external documentation. Both our proposed system and the baseline are based on the same underlying LLM, GPT-4o.

3.5.2 Scenarios settings

To ensure that participants engage meaningfully within the scope of this study, the experiment introduces three structured scenarios. These scenarios are designed to stimulate in-depth conversations that reflect real-world workplace safety challenges in the construction industry. However, participants maintain autonomy over the specific direction and content of their conversation, even though their personas are the individuals within the scenario:

> **Scenario 1:** Safety Concerns and Lack of Protective Equipment. *"You are a 27-year-old construction worker operating heavy machinery on-site. Lately, you've noticed that some of your coworkers are* *not wearing personal protective equipment (PPE) due to a lack of proper gear provided by the company. When you brought this up, your supervisor dismissed your concerns, saying, "We've never had a major accident. Just be careful." You're worried about safety violations and potential hazards but are unsure how to address them without risking your job."*

**Scenario 2:** Mental Health Struggles After a Workplace Injury. *"You are a 45-year-old carpenter who recently suffered a fall at work, injuring your shoulder. Although your doctor recommended rest, you feel pressured to return early due to financial concerns. Your coworkers tell you, "Tough it out—we all get hurt on the job." You're experiencing chronic pain, stress, and fear of taking time off."*

**Scenario 3:** Burnout and Lack of Job Recognition. *"You are a 50-year-old crane operator who has worked in construction for 25 years. Lately, you feel burnt out and unappreciated, as younger workers are getting promoted while you are assigned repetitive, physically demanding tasks. You don't feel like you have much say in your work, but you have a family to support and can't afford to leave. You're struggling with motivation, stress, and job dissatisfaction."*

3.5.3    User Study Procedure

In this study, we recruited 12 participants, comprising 8 males and 4 females. Each participant had a one-on-one session. Prior to the study, participants received a briefing outlining the experimental procedures, the functionalities of the system, and the specific roles of the agents within the proposed system.

At the beginning, the participants were randomly assigned to one of the scenarios detailed in Section 3.5.2. Within their assigned scenario, each participant interacted with both the proposed system and the baseline system, seeking for insights and useful solutions during the interaction. To mitigate potential learning effects, the order of system exposure was randomized. After completing the interaction with the first system, participants immediately completed a survey assessing their experience. They then proceeded to interact with the second system and subsequently filled out the same survey, then the user study concluded.

Surveys utilized in this study comprise four sections: (1) System Usability Scale (SUS): This section includes the standard 10-item SUS scales (Brooke 1996). (2) Self-Determination: This section was measured by the basic psychological needs satisfaction survey (La Guardia et al. 2000): autonomy (three items), competence (three items), relatedness (three items). (3) Perceived Social Presence and Trust: This section was measured by three questions includes social presence question, answer-trust question and adoption intent question. (4) Open-ended Question: Comments for overall experimental experience.

## 4. RESULTS AND ANALYSIS

### 4.1   SUS Results and Analysis

Figure 3 presents the comparison of SUS rating distribution between the baseline system and our proposed system. For each question, the value in each category cell represents the number of participants who provided that rating. Participants consistently found our system has a higher usability than the baseline system. According to the standard SUS calculation (Brooke 1996) we normalize the raw SUS score to a 100-point scale. Figure 4 presents a comparison of the normalized SUS scores between the baseline system and our proposed system. The statistical analysis reveals that users rated our system (Mean = 84.58, SD = 8.95) significantly higher than the baseline system (Mean = 71.88, SD = 16.46) in terms of normalized SUS score ($p = .0083$).

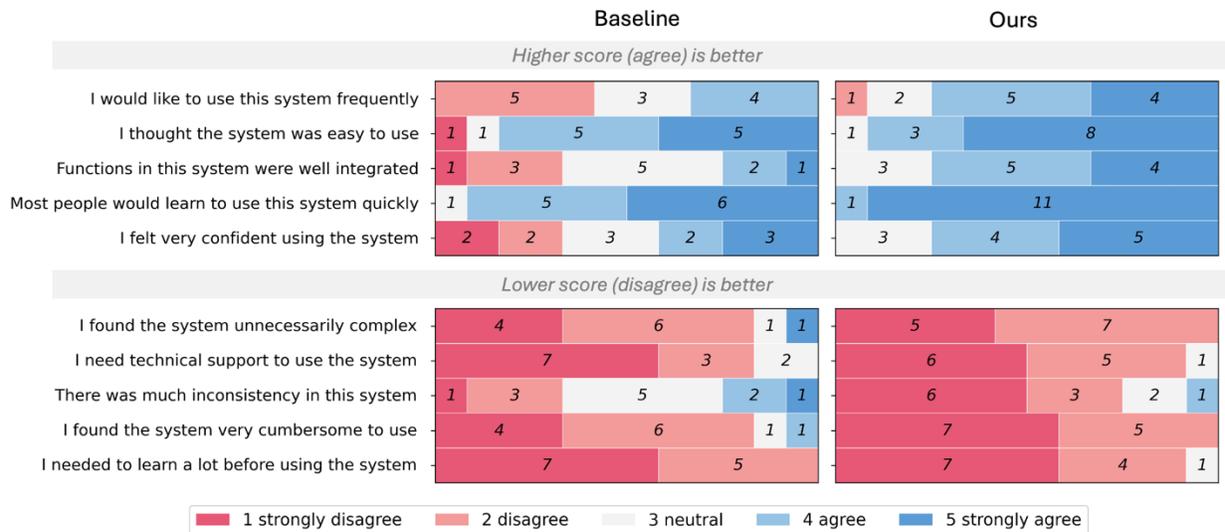

Figure 3: System Usability Scale (SUS) rating distribution: Baseline vs. Our system

According to the SUS Sauro-Lewis curved grading scale (Lewis and Sauro 2018), our multi-agent system falls within the 'A Grade' range. This suggests that users found the multi-agent system more intuitive and effective in supporting workers' problem-solving in these scenarios, making it well-suited for practical deployment. This result aligns with the responses in the open-ended question, where 11 out of 12 participants indicated a preference for our proposed system. Some quotations from participants' comments are as follows: P3 commented, "*I don't think the baseline gave me any specific suggestions that I can take directly. The response from the baseline is too general, and I don't feel any humanity.*" P8 commented, "*The multi agent system is quite helpful. The baseline can also reach the goal but requires much more energy, its convenience is worse that the multi-agent system in this circumstance.*"

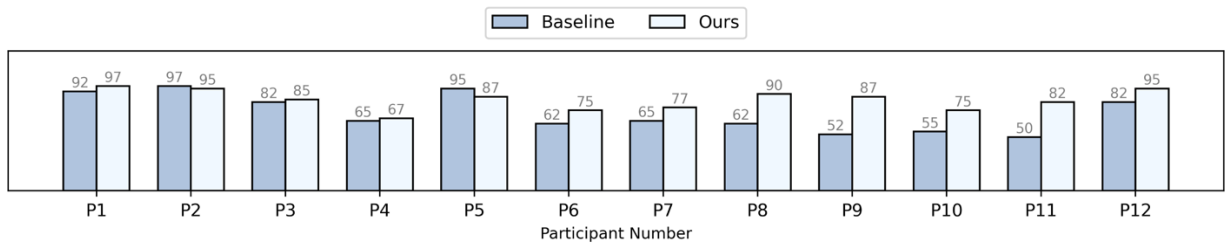

Figure 4: Comparison of normalized SUS (%) scores between baseline and our proposed system

## 4.2  Self-Determination Results and Analysis

Figure 5 presents a comparative analysis of user ratings for the baseline system and our Multi-Agent Chat system across three key dimensions: *Autonomy*, *Competence*, and *Relatedness* (La Guardia et al. 2000). The results indicate that our proposed system consistently outperforms the baseline in all three aspects. For *Autonomy*, users rated our system higher (Mean = 4.14, SD = 0.65) compared to the baseline (Mean = 3.50, SD = 0.89), with a marginally significant difference ($p = 0.066$). In terms of perceived competence, our system received a significantly higher rating (Mean = 3.86, SD = 0.59) than the baseline (Mean = 3.25, SD = 0.77), achieving statistical significance ($p = 0.048$). The most pronounced improvement is observed in Relatedness, where our system scored substantially higher (Mean = 3.83, SD = 0.57) than the baseline (Mean = 2.08, SD = 0.64), with a highly significant difference ($p < 0.001$). These findings suggest that the multi-agent chat system can foster greater user autonomy, enhance perceived competence, and significantly improve social connectedness for workers.

Participants reported greater autonomy while interacting with the multi-agent system, suggesting that it provided more flexibility and control over the conversation. The system also enhanced perceived competence, as users felt more confident in understanding and utilizing the system's responses. Additionally, higher relatedness scores indicate that users experienced a stronger sense of connection and engagement with the system. These findings highlight the potential of multi-agent AI systems in fostering workers' well-being through a more engaging and supportive interaction. This result aligns with the users' comments, some quotations from participants are as follows: P1 commented, "*I like the multi-agent system because I could quickly process the answers, even for users without relevant knowledge. In contrast, the baseline provides a long list of responses that require more logical processing.*" P9 commented, "*The baseline can't give me emotional support, I feel it's only analyzing my problems without supporting me. In the multi-agent system, I feel I have been companied with my colleagues.*"

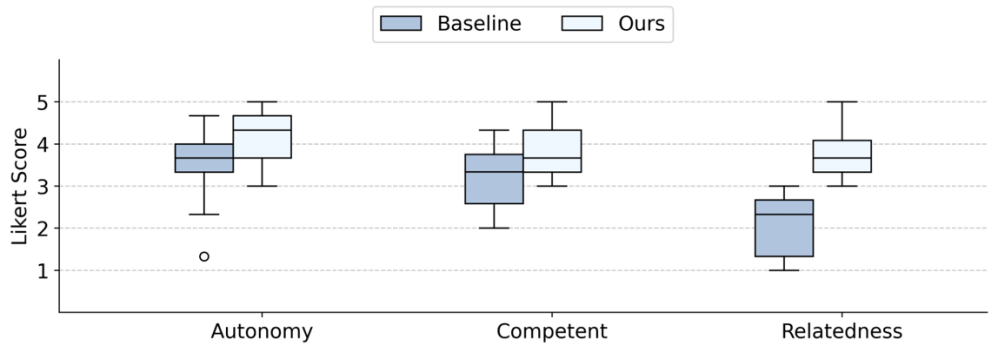

Figure 5: Comparison of Self-Determination scores between baseline and our proposed system

### 4.3 Perceived Social Presence and Trust Results and Analysis

#### 4.3.1 Reliability analysis of the scale

In this study, we examined the reliability of a three-item scale designed to measure Perceived Social Presence and Trust in the proposed system and the baseline system. The scale consists of three items (Q1-Q3) as shown in Figure 6. The three items aim to capture users perceived social presence (i.e., whether the AI feels human-like), trust (i.e., whether the answer is practical), in the system's responses, and intend to adopt the system for future use.

To assess the internal consistency of this scale, we conducted Cronbach's Alpha reliability analysis using data collected from the three questions in perceived social presence and trust scale. Cronbach's Alpha values above 0.7 are generally considered acceptable, while values above 0.8 indicate good reliability, and values above 0.9 indicate excellent reliability (Tavakol and Dennick 2011). The calculated Cronbach's Alpha results shows the proposed system has an 'Excellent reliability' ($\alpha = 0.911$) and the baseline system has a 'Good reliability' ($\alpha = 0.817$).

The results demonstrate that the three items in our Perceived Social Presence and Trust scale exhibit strong internal consistency, suggesting that they collectively measure a coherent underlying construct.

#### 4.3.2 Results analysis

Figure 6 presents the comparison of the 3-item scale rating distribution between the baseline system and our proposed system. The results indicate that the multi-agent system received significantly higher ratings across all three dimensions. For Q1, users rated our system higher (Mean = 3.33, SD = 0.62) compared to the baseline (Mean = 1.42, SD = 0.49), with a highly significant difference ($p < .001$). In terms of Q2, our system received a higher rating (Mean = 4.00, SD = 0.82) than the baseline (Mean = 2.92, SD = 1.04), achieving significance ($p = .013$). For Q3, our system scored substantially higher (Mean = 4.00, SD = 0.82) than the baseline (Mean = 2.75, SD = 0.83), with a very significant difference ($p = .002$).

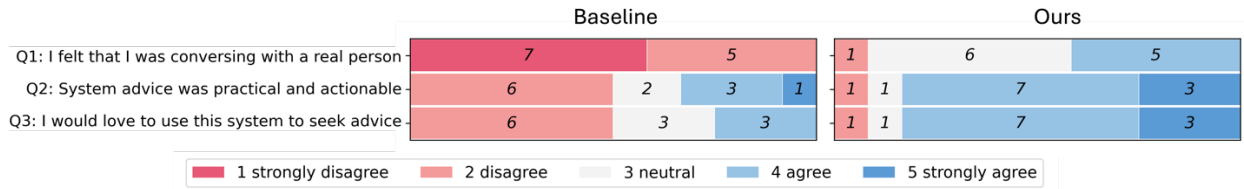

Figure 6: Perceived Social Presence and Trust rating distribution: Baseline vs. Our system

The results demonstrate that it was perceived as more natural and human-like in conversations, provided more useful advice, and elicited a stronger willingness among users to adopt it for future use. Based on user feedback, we found that the proposed system's agent identities, professional profiles, and domain-specific expertise played a crucial role in driving higher ratings across these questions. These elements enhanced user engagement, trust, and the overall perceived social presence and trust of the system, contributing to a more natural and informative interaction experience. Some quotations from participants' comments are as follows: P7 commented, "*The interactions with multi-agent are closer to consulting with real experts. their reply contains more care, making me feel more comfortable*." P11 commented, "*I think multiagent can give me more practical suggestions and also from different perspectives.*" P12 commented, "*I prefer the multi-agent chat, it was easier to take in the information. it was a good experience. It is really cool that each chat bots had their own personality and background. it made them feel more real and human*."

## 5. CONCLUSIONS AND DISCUSSION

This study set out to design and evaluate a conversational multi-agent AI system tailored to the mental health and safety challenges of construction workers. Through systematic work includes: interface design, system development and comparative user study, we demonstrated how the proposed system can serve as a tool to leverage the power of LLM, even for non-experts, since the system enables users to easily customize the LLM and external knowledge integration. Through the user study, we demonstrated the proposed system's exceptional performance in terms of usability, as well as its ability to enhance workers' psychological needs, social presence and trust than a generic baseline chatbot.

Beyond simply answering queries, the multi-agent framework fosters autonomy, competence, and relatedness—key contributors to human basic psychological needs. This approach underscores the potential of conversational AI to provide timely, relevant, and empathetic support in construction industry. As a result, construction workers not only gain credible guidance on safety and mental health but also benefit from a more human-like, empowering, and socially engaging interaction. This work lays the groundwork for broader adoption of human-AI-interaction for worker support in construction industry, setting a promising direction for future research and practical deployment in the construction sector and beyond. While this study demonstrates the system's potential, the small-sized participant group and predefined scenarios only represent our initial exploration. Future work will include studying more diverse user groups such as real construction workers and exploring scenarios tailored to each different individual.